# Efficient and Scalable GaInAs Thermophotovoltaic Devices

Eric J. Tervo,[1] Ryan M. France,[1] Daniel J. Friedman,[1] Madhan K. Arulanandam,[1,2] Richard R. King,[2] Tarun C. Narayan,[3] Cecilia Luciano,[3] Dustin P. Nizamian,[3] Benjamin A. Johnson,[3] Alexandra R. Young,[3] Leah Y. Kuritzky,[3] Emmett E. Perl,[3] Moritz Limpinsel,[3] Brendan M. Kayes,[3] Andrew J. Ponec,[3] David M. Bierman,[3] Justin A. Briggs,[3] and Myles A. Steiner[1,*]

[1]*National Renewable Energy Laboratory, Golden, CO, 80401, USA*

[2]*School of Electrical, Computer, and Energy Engineering, Arizona State University, Tempe, AZ, 85281, USA*

[3]*Antora Energy, Inc., Sunnyvale, CA, 94089, USA*

[*]myles.steiner@nrel.gov

Thermophotovoltaics are promising solid-state energy converters for a variety of applications such as grid-scale energy storage, concentrating solar-thermal power, and waste heat recovery. Here, we report the design, fabrication, and testing of large area (0.8 cm$^2$), scalable, single junction 0.74-eV GaInAs thermophotovoltaic devices reaching an efficiency of 38.8±2.0% and an electrical power density of 3.78 W/cm$^2$ at an emitter temperature of 1850°C. Reaching such a high emitter temperature and power density without sacrificing efficiency is a direct result of combining good spectral management with a highly optimized cell architecture, excellent material quality, and very low series resistance. Importantly, fabrication of 12 high-performing devices on a two-inch wafer is shown to be repeatable, and the cell design can be readily transferred to commercial epitaxy on even larger wafers. Further improvements in efficiency can be obtained by using a multijunction architecture, and early results for a two-junction 0.84-eV GaInPAs / 0.74-eV GaInAs device illustrate this promise.

# INTRODUCTION

Thermophotovoltaic (TPV) devices are solid-state converters of heat to electricity that operate similarly to solar photovoltaics. Instead of capturing photons from the sun, however, TPV cells absorb and convert infrared photons radiated from a hot thermal emitter.[1-7] This allows TPVs to be driven by a variety of heat sources and to be used in many different applications, including solar-thermal energy conversion,[8-11] waste-heat recovery from industrial processes,[12-14] combustion-fired electricity generation,[15-18] and large-scale thermal energy storage.[19-22]

TPV devices have two important characteristics that are distinct from typical solar photovoltaic cells. First, they must manage and utilize a very broad spectrum of light radiating from thermal emitters at temperatures lower than that of the sun. A significant portion of the blackbody radiation spectrum will lie at energies below the TPV cell's bandgap, and strategies must be employed to prevent the emission of these low-energy photons or reflect them back to the emitter.[2, 3, 7] Parasitic absorption of either below-bandgap-energy photons or of above-bandgap-energy photons in undesired locations (e.g. a back contact layer), as shown schematically in Fig. 1A, is one of the primary loss mechanisms that degrades the spectral efficiency $SE$ or internal quantum efficiency $IQE$ (Methods) of TPV devices.[5, 7] Second, TPV cells tend to operate at much higher current and power densities than solar photovoltaics, because the radiation intensity from an emitter with a high view factor to the cell exceeds that from the sun even at relatively low emitter temperatures.[2, 3] This requires that TPV cells have a very low series resistance to reduce ohmic losses (which degrade the fill factor $FF$) in addition to avoiding nonradiative recombination losses (which degrade the voltage factor $VF$) (Methods).[5, 23] These loss pathways are also shown in Fig. 1A.

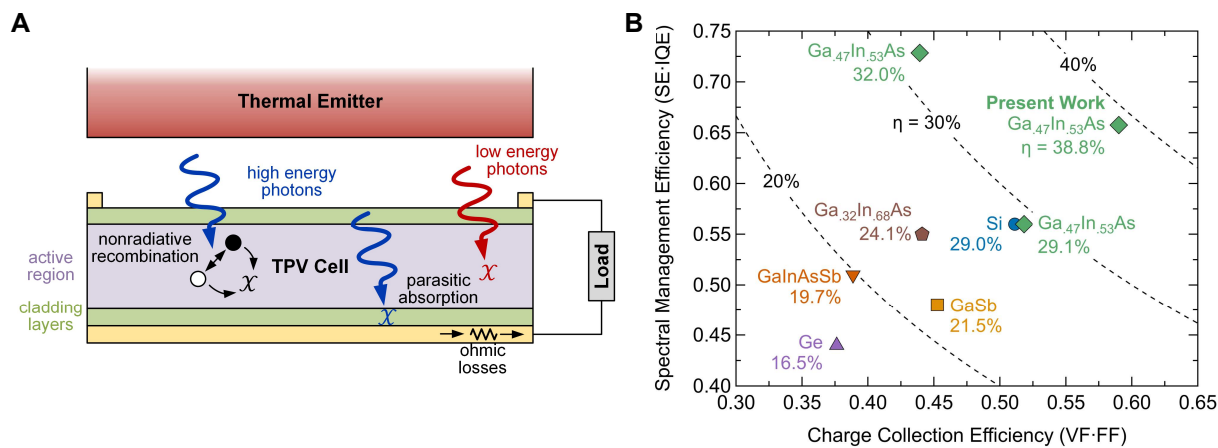

**Figure 1. Loss mechanisms and efficiencies of TPV devices.**
(A) Schematic of a TPV emitter and cell illustrating the primary loss mechanisms that contribute to spectral management inefficiency (parasitic absorption of above- or below-bandgap photons as well as thermalization losses, not shown) and charge collection inefficiency (nonradiative recombination and ohmic losses).
(B) Historical and present experimental single-junction TPV efficiencies (given by point labels). Dashed lines are lines of constant TPV efficiency. Improvements to TPV performance require minimizing all loss mechanisms illustrated in (A) to achieve both high spectral management efficiency and high charge collection efficiency. Past leading efficiency measurements compiled by Burger *et al.*[6] include Ge,[24] GaInAsSb,[25] GaSb,[26] $Ga_{0.32}In_{0.68}As$,[27] Si,[28] and $G_{0.47}In_{0.53}As$[29, 30] cells.

Researchers have developed single-junction TPV devices from a variety of crystalline semiconductor materials, and the best measured TPV efficiencies are shown in Fig. 1B.[6, 24-30] Much of this work has focused primarily on improving spectral efficiency by reducing absorption of sub-bandgap photons. For example, the 24.1% $Ga_{0.32}In_{0.68}As$ device in Fig. 1B used both a selective emitter to reduce emission of very low-energy photons and a front surface filter to reflect photons of energy just below the bandgap back to the emitter,[25] while the ~29% Si and $Ga_{0.47}In_{0.53}As$ devices used highly reflective mirrors behind the cell to reflect most sub-bandgap photons back to the emitter.[28-30] However, spectral efficiency is just one (albeit important) component of overall TPV efficiency:[6]

$$\eta_{TPV} = \underbrace{(SE \cdot IQE)}_{\text{Spectral Management Efficiency}} \cdot \underbrace{(VF \cdot FF)}_{\text{Charge Collection Efficiency}} \quad (1)$$

Fig. 1B breaks down TPV efficiencies according to the two factors given in Eq. 1 and explained in greater detail in the Methods, and it also highlights the necessity of optimizing both the spectral management and charge collection characteristics. Our single-junction 0.74-eV GaInAs cells presented in this work reach nearly 40% efficiency by minimizing all loss mechanisms in a TPV device. Specifically, our cells achieve a spectral management efficiency of 65.8% and a charge collection efficiency of 59.0%, which are 89.4% and 74.3% of the respective radiative limits (Fig. S3).

A key aspect of moving TPV technology from the laboratory to practical applications is ensuring scalability and reproducibility of the cells. The best recent $Ga_{0.47}In_{0.53}As$ cells,[29, 30] for example, had active areas ≤0.1 cm$^2$, and no data were reported on the reproducibility of these cells. Scaling up to large areas is particularly problematic, because series resistance increases significantly as the cell area increases,[31] but larger cells facilitate large module manufacturing. Recently, two-junction tandem TPV cells grown on lattice-mismatched substrates via a metamorphic buffer have been shown to reach efficiencies as high as 41.1%,[32, 33] but their structure is much more complex than single junction cells grown on lattice-matched substrates and could therefore impose significant manufacturing challenges or costs. To address these concerns, we show that our large-area (0.8 cm$^2$) cells can be reliably and repeatably grown on large (2-in) wafers while still maintaining very low series resistance.

**RESULTS AND DISCUSSION**

**Cell Fabrication**

Our TPV cells are grown by metalorganic vapor phase epitaxy (Methods), which produces very high-quality III-V semiconductor material and is regularly used in the commercial sector to produce photovoltaic cells for spacecraft applications. Beginning with a 2-in InP substrate, all epitaxial layers are grown with the same lattice constant to minimize defects, and significant care is taken with the cell architecture (Fig. 2A) to minimize parasitic absorption and maximize efficiency. For example, the window layer (which serves to passivate the front side of the GaInAs absorber) is made as thin as possible to reduce sub-bandgap free-carrier absorption, and the back contact layer (which serves to make ohmic contact with the back metal reflector) is alloyed with Al to raise the bandgap and reduce parasitic absorption of useful photons that could be reflected back to the absorber layer.

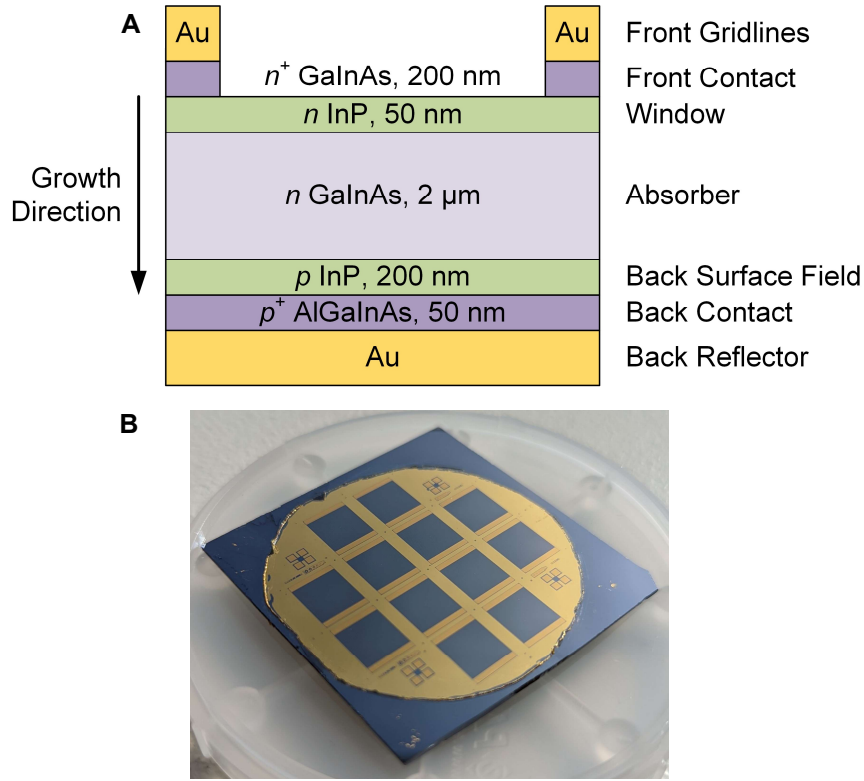

**Figure 2. Single-junction 0.74-eV GaInAs TPV cells.**
(A) Schematic of the single rear-heterojunction TPV cell structure. Layers other than the absorber are kept as thin as possible to reduce parasitic above- and below-bandgap absorption.
(B) Photograph of 12 TPV cells fabricated from a two-inch wafer and bonded to an inert silicon wafer for mechanical handling and stability.

The cells are grown inverted, such that the front contact layer is grown first on the InP substrate, and the back contact layer is grown last. This allows the back reflector to be deposited first after epitaxial growth, followed by bonding the entire structure to a carrier material (Si), after which the substrate is chemically etched away to leave just the thin epitaxial material. The individual cells are then fabricated with standard photolithography, electroplating, and wet chemical etching techniques (Methods). It is important to note that the electroplated back reflector and front gridlines are very thick (4 μm and >6 μm, respectively) to reduce series resistance. Inverted growth and processing is a common practice for III-V photovoltaic cells, because it enables the creation of lightweight, flexible, thin-film cells[34, 35] and allows for substrate reuse (and therefore significant reduction of costs) with methods such as epitaxial liftoff and spalling.[36-40] A photograph of 12 TPV cells fabricated from a 2-in wafer is shown in Fig. 2B, where each cell has dimensions of 0.8 cm by 1 cm and an aperture area (excluding busbars) of 0.64 cm². We do not deposit an antireflection coating on these cells; such a coating would increase the photocurrent and cause series resistance to play a limiting role at lower emitter temperatures, as discussed later. An antireflection coating would, however, increase the electrical power density, which may be an advantageous tradeoff depending on the specific application.

**Cell Characterization**

As discussed in the introduction, a variety of TPV cell electrical and optical characteristics have a strong effect on the TPV efficiency. One of the most fundamental measures of material quality and spectral management is the $IQE$, calculated from the measured external quantum efficiency and the measured reflectance as shown in Fig. 3A. These data are for a 0.74-eV GaInAs cell with the same structure as the devices shown in Fig. 2 but with shorter and more widely spaced front gridlines, because the tall, dense gridlines on a TPV cell result in light scattering that makes specular reflectance measurements inaccurate. We obtain a high, broad $IQE$ which indicates that there is very little parasitic absorption of useful above-bandgap photons in undesired regions of the cell (such as the window or back contact layers) and indicates that electron-hole pairs generated in the absorber layer are efficiently separated and collected by the junction. The apparent shift in bandgap between the external and internal quantum efficiencies is a known artifact in high-reflectance samples, and the external quantum efficiency is the true indicator of bandgap.[41]

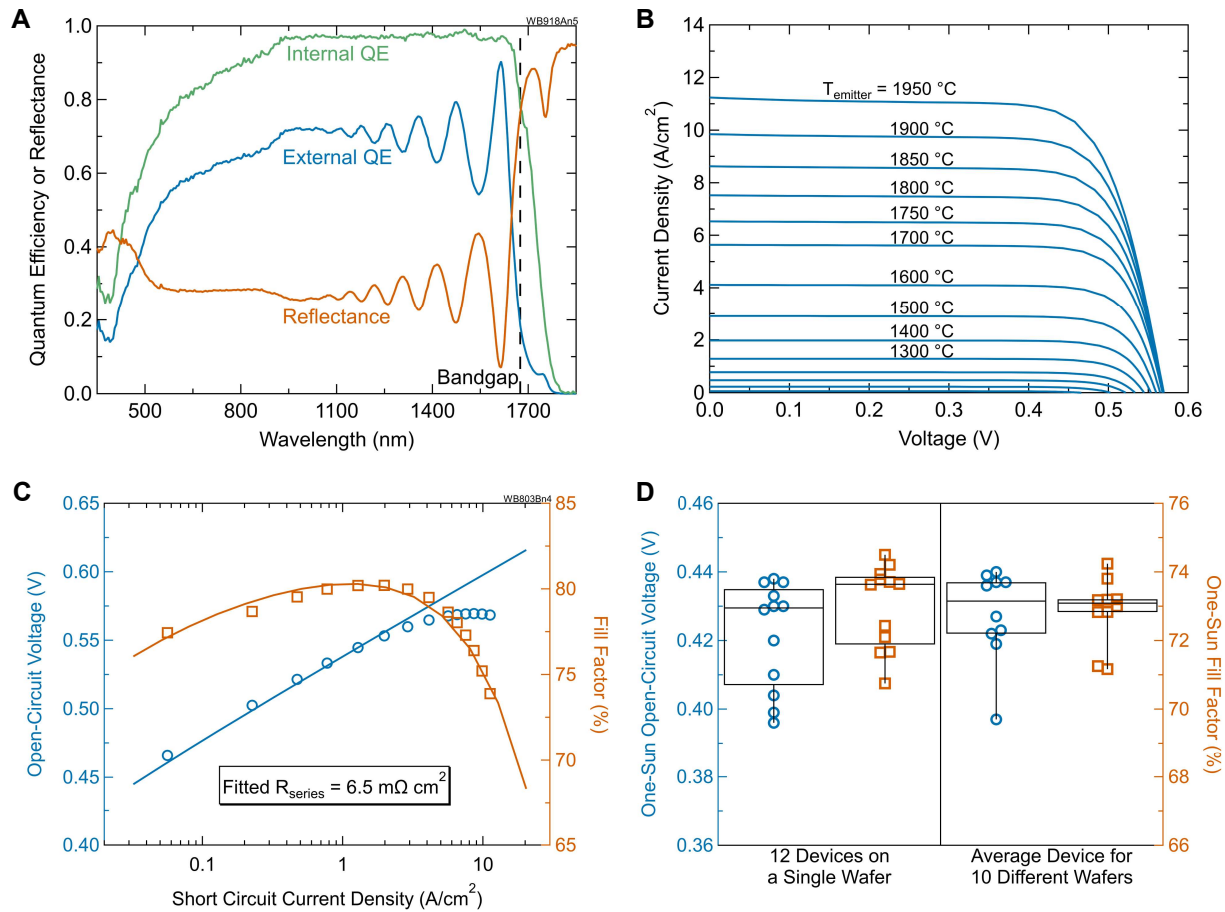

**Figure 3. Single-junction TPV cell electrical characteristics.**
(A) Measured reflectance and external quantum efficiency and calculated internal quantum efficiency for a single-junction 0.74-eV GaInAs test cell with the same structure as our TPV cells.
(B) Current-voltage measurements of a TPV cell illuminated by a thermal emitter with varying temperature.
(C) Open-circuit voltage (blue circles) and fill factor (orange squares) extracted from (B). Solid lines are data fit to a generalized optoelectronic model that provides the series resistance.[42]
(D) Measured one-sun open-circuit voltage (blue circles) and fill factor (orange squares) for many devices and growths. The horizontal lines indicate the 75th percentile, median, and 25th percentile. The left half shows data for 12 TPV cells fabricated on a single 2-in wafer, and the right half shows the average performance of all devices on a single wafer for 10 different wafers. For all data, the relative tight grouping of voltage and fill factor indicate reproducibility across a single and multiple wafers.

Additional information about cell electrical characteristics is obtained from current-voltage measurements under illumination. Fig. 3B shows current-voltage data obtained from a 0.74-eV GaInAs cell while undergoing TPV efficiency testing, and Fig. S1 shows current-voltage data obtained under simulated AM1.5Direct illumination (Methods), which is a useful standard reference spectrum for cell development and comparison to other devices. As the temperature of the emitter increases, the radiation incident on the cell increases both in intensity and in photon energy, which causes the increase in photocurrent shown in Fig. 3B. The short-circuit current, fill factor, and open-circuit voltage are extracted from these data and plotted in Fig. 3C for the cell undergoing efficiency testing and noted on Fig. S1 for the cell under AM1.5Direct illumination. Our cell has an excellent open-circuit voltage under one sun of 0.443 V (with short-circuit current of 33.9 mA/cm$^2$), corresponding to a bandgap-voltage offset ($E_g/q - V_{oc}$, where $E_g$ is the bandgap energy, $q$ is the electron charge, and $V_{oc}$ is the open-circuit voltage) of 0.297 V which is close to the radiative limit of 0.285 V for this bandgap.[43, 44] The open-circuit voltage increases as expected with increasing radiation intensity (Fig. 3C). The departure from the logarithmic trend results from the cell temperature increasing slightly as the emitter temperature increases (Fig. S2) due to limitations in the thermal management system used here (Methods). The *FF* is also high at one sun (74.0%, Fig. S1) and at low emitter temperatures, and it remains high even at very high current densities of multiple A/cm$^2$ (Fig. 3C). The impressive *FF* at high current densities largely results from efforts to reduce series resistance through the use of a thick back reflector and front gridlines, as discussed in the "Cell Fabrication" section, as well as a tight pitch of the front gridlines (50 μm pitch and 10 μm width). To determine the series resistance of this cell, an optoelectronic model[42] was fit to these data (Methods), giving a very low series resistance of 6.5 mΩ cm$^2$.

Although Figs. 3B, 3C, and later figures show data for one of our best-performing cells, these characteristics are repeatable across many devices on a single wafer and across multiple growths on different wafers. Fig. 3D shows open-circuit voltage and *FF* data for a number of different devices and growths under simulated AM1.5Direct illumination, all of which have the same structure as indicated in Fig. 2A except for slightly different thicknesses of the window and back contact layers and/or different front gridline pitch, in some cases. The ability to consistently achieve open-circuit voltages >0.40 V and *FF* >70% indicates that our TPV cell architecture can be reliably grown and fabricated, even using a small research-scale growth reactor. Transferring this design to larger wafers in commercial epitaxy facilities that are optimized for uniformity and repeatability should further improve the consistency of these devices.

Our TPV cells are designed to be used with emitters that have high broadband emittance similar to a blackbody, which means that the cells must reflect as much sub-bandgap light back to the emitter as possible. We use our thin-film cell as a selective filter that is transparent to sub-bandgap light and place a back reflector behind the cell to reflect this light, which is the same strategy as that employed in recent studies of 0.74-eV GaInAs devices[29, 30] included in Fig. 1B. The hemispherical reflectance is measured via Fourier transform infrared reflectance spectroscopy and modeled with the standard transfer matrix method[45] (Methods), and these results are shown in Fig. 4A along with the normalized blackbody emission spectrum at 1850°C for reference. The cell achieves a high sub-bandgap reflectance of 94.7% when weighted to this blackbody spectrum from 1.72 μm (20 meV below the nominal bandgap energy to avoid the band edge) to 15 μm. The sources of sub-bandgap absorption are revealed with our model by calculating the spectral absorption in each layer of the cell, shown in Fig. 4B. Most of this absorption occurs in the Au back reflector, which results from waveguide modes that exist at the semiconductor-metal interface.[30, 46] At longer wavelengths, free-carrier absorption in some of the cell layers becomes more significant, but the radiation intensity also falls off significantly at longer wavelengths (Fig. 4A), making this less important.

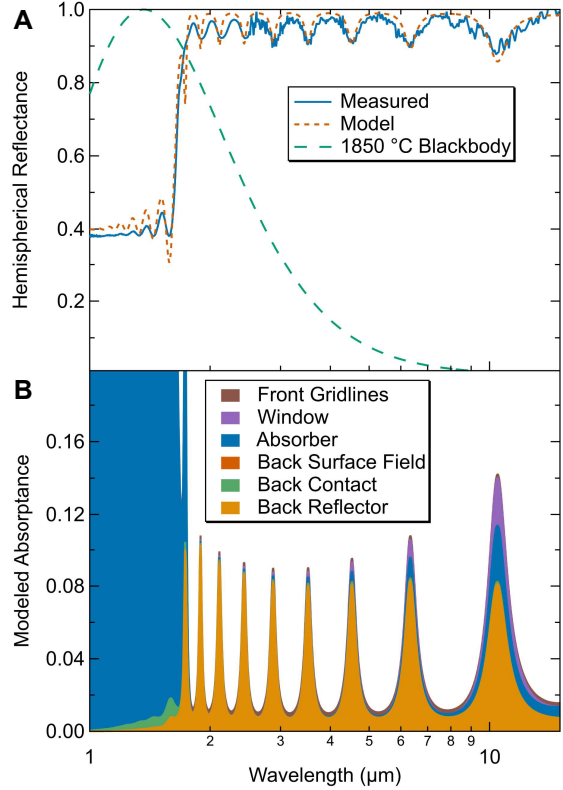

**Figure 4. Single-junction TPV cell optical characteristics.**
(A) Measured and modeled hemispherical reflectance of a TPV cell under Lambertian illumination. Also shown is the normalized 1850°C blackbody spectrum, which indicates that the amount of sub-bandgap energy at wavelengths longer than 15 μm is negligible.
(B) Modeled absorptance in each layer for the same case as (A). Parasitic absorption for wavelengths longer than the bandgap occurs primarily in the Au back reflector with a small amount of free-carrier absorption in the window and absorber.

Our measured and modeled weighted sub-bandgap reflectance of 94.7% is approximately the same as that reported for a 0.74-eV GaInAs device by Omair *et al.* (94.6%), and they also have a similarly high $IQE$.[29] Nevertheless, our spectral management efficiency (65.8%) is about ten points higher than theirs (56.0%) as shown in Fig. 1B. This is because our cell was tested with a much higher emitter temperature (1850°C compared to 1207°C) and not because of significant long-wavelength optical differences between the cells. For a particular bandgap energy, higher emitter temperatures have a higher portion of the thermal blackbody spectrum above the bandgap. This means that, for a particular bandgap energy, sub-bandgap absorption causes a greater penalty with lower temperature emitters than with higher emitter temperatures. In other words, lower emitter temperatures require higher sub-bandgap reflectance to reach the same spectral management efficiency. Achieving high efficiency at high temperatures, however, also requires high charge collection efficiencies and especially a high $FF$ enabled by low series resistance. Due to our low series resistance of 6.5 mΩ cm$^2$, our $FF$ does not begin to decrease until photocurrents greater than 2 A/cm$^2$ are reached (Fig. 3C), allowing us to reach very high emitter temperatures without a substantial penalty to TPV efficiency. Omair *et al.*'s higher series resistance of 43 mΩ cm$^2$ means that *FF* will begin to decrease and limit the TPV efficiency at much lower photocurrents and hence emitter temperatures.

A similar comparison can be made to the recent results of a 0.74-eV GaInAs device by Fan *et al.*, which achieved an impressive 98.9% weighted sub-bandgap reflectance at an emitter temperature of 1182°C due to their air-bridge architecture.[30] This very high reflectance allowed them to reach a higher spectral management efficiency of 72.9% (Fig. 1B) despite operating at a much lower emitter temperature than our device. Once again, however, the TPV efficiency at high temperatures is limited due to the high series resistance of 26 mΩ cm$^2$ that causes *FF* to decrease above photocurrents of about 0.2 A/cm$^2$. As a result, when our spectral management efficiencies are normalized to the radiative limit (Methods), we obtain similar values (89.2% for Fan *et al.* and 89.4% for our cell), as shown in Fig. S3. These two comparisons to prior work further demonstrate the importance of optimizing TPV efficiency through progress on multiple fronts, including both electrical and optical characteristics.

**TPV Efficiency**

The TPV efficiency in this study is determined calorimetrically through simultaneous measurement of the electrical power produced by the cell, $P$, and the heat dissipated from the cell, $Q$, in a custom test facility while under illumination from a heated graphite emitter through a water-cooled copper aperture (Methods).[47, 48] We emphasize that the use of a heated graphite emitter provides a realistic thermal spectrum similar to what would be seen in a practical application. It is also important to note that this measurement approach does not rely on a separate measurement of the cell's reflectance to infer efficiency. This allows the efficiency to be directly determined from measured quantities as[49, 50]

$$\eta_{TPV} = \frac{P}{P + Q} \qquad (2)$$

Efficiency results are given in Fig. 5 for the 0.74-eV GaInAs cell described previously with emitter temperatures between 1200–2000°C. This cell reaches a maximum efficiency of 38.8±2.0% at an emitter temperature of 1850°C. The electrical power density produced at this temperature is 3.78 W/cm$^2$, which is very large in comparison to most experimental TPV research.[5] The ability to reach such a high emitter temperature and power density without sacrificing efficiency is a direct result of the combination of excellent optical and electrical characteristics of our cell. Above 1850°C, at which point the photocurrent is 8.66 A/cm$^2$, continued decreases in *FF* (Fig. 3C) due to series resistance cause the efficiency to begin falling. Our efficiency model (Methods) shows good general agreement with the trends in efficiency, although it overpredicts the experimental data for lower emitter temperatures. This may be caused by possible discrepancies between the directional-hemispherical reflectance measurements we made and the true bihemispherical reflectance of the cell, or it may be caused by possible parasitic heat paths in the measurement apparatus.

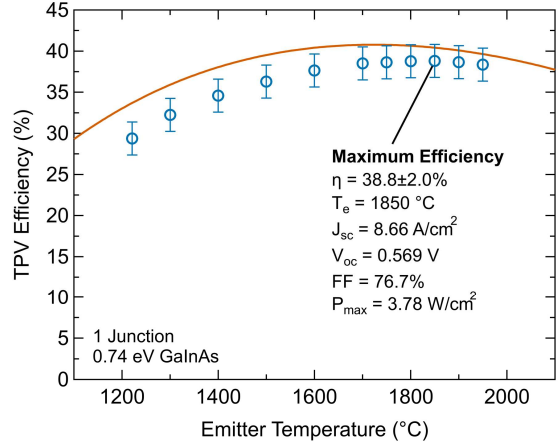

**Figure 5. Single-junction TPV efficiency.**
Measured (points) and modeled (solid line) heat-to-electricity conversion efficiency of a single-junction 0.74-eV GaInAs cell with 0.8 cm² total and 0.64 cm² aperture areas. The efficiency begins to decrease above 1850°C due to series resistance and high current densities.

**Potential for Efficiency and Power Gains with Tandem Cells**

One promising pathway to reach even higher efficiency is with the use of a tandem TPV cell with two or more junctions. This provides two key benefits: first, a tandem consisting of materials with different bandgap energies allows incident light to be separated by energy and absorbed with lower thermalization losses; and second, series-connected sub-cells in a tandem device operate at half the current density and more than twice the voltage of a single-junction device with the same lowest bandgap energy, which reduces the need for very low series resistance and allows for efficient operation at higher emitter temperatures. Reducing cell current is a potent advantage, since halving the current density reduces the fractional power loss due to resistance by a factor of four.

To demonstrate this potential, we provide here results for preliminary two-junction tandem 0.84-eV GaInPAs / 0.74-eV GaInAs cells, whose structure is shown in Fig. S4. As with the single-junction cells, these are grown lattice-matched to InP wafers (Methods), which provides good material quality and makes the design potentially more scalable than tandem cells that require a metamorphic buffer layer to change the lattice constant between the substrate and the cell or between sub-cells.[32, 33] Key to the design of any series-connected tandem is the interconnection between sub-cells. Here we use a $p^+$ GaAsSb / $n^+$ InP tunnel junction, where the GaAsSb was used on the p-type side instead of GaInAs or InP because of the ability to heavily dope with carbon.

The electrical characteristics of this tandem cell are shown in Fig. S5, and the optical characteristics are shown in Fig. S6. A broader quantum efficiency is obtained with the tandem cell than with the single-junction cell, as expected. It should be noted that the measured *IQE* for the tandem cell may be lower than its true value because the measured cell had tall grids that are known to scatter light away from the detector. The primary advantages of the tandem cell are clearly shown by the open-circuit voltage and *FF* data in Fig. S5C. The open-circuit voltage is >1.12 V for even the lowest emitter temperatures and photocurrents and is more than twice the voltage of the single junction cell for all current densities, which results from the summation of the sub-cell voltages in series. The sharp minimum in the *FF* at 1700°C indicates that the two sub-cells are current-matched at that temperature[51] as a result of slightly unoptimized cell thickness or bandgap for the actual operation at temperatures >1800°C where the highest efficiencies were measured. Further adjustments to cell thicknesses or bandgaps could result in even higher efficiencies. The lowest *FF* is >78%, despite a higher series resistance of ~10 mΩ cm$^2$ for the two-junction cells compared to 6.5 mΩ cm$^2$ for the single-junction cells. The high *FF* is made possible by the lower current densities. The sub-bandgap reflectance of the tandem cell weighted to a thermal emitter temperature of 1900°C is 92.7%. Modeling indicates that this lower reflectance is largely due to free-carrier absorption in the highly-doped tunnel junction layers, as shown in Fig. S6B, which can potentially be improved through doping and thickness optimizations.

Efficiency results for this tandem device are obtained from the same test facility as that used for the single-junction device and are shown in Fig. 6. The cell reaches a maximum efficiency of 36.8±2.0% at the highest emitter temperature of 1900°C. Even though the maximum efficiency of the tandem cell is lower than the single-junction cell, three characteristics stand out that demonstrate the potential of tandems to reach higher efficiencies. First, the power density of 5.65 W/cm$^2$ is higher than that of the single-junction cell at its maximum efficiency point but with a lower photocurrent of 5.92 A/cm$^2$, which illustrates ability of a tandem to provide high power without significant detrimental effects of series resistance. Second, the fact that the efficiency is still rising at the highest emitter temperature tested illustrates that tandems function well with high emitter temperatures relative to their lowest bandgap energy. Third, as described previously, operation with higher emitter temperatures will reduce the penalties associated with lower sub-bandgap reflectance. These characteristics of tandem cells indicate that further optimization of the cell, such as reducing absorption in the tunnel junction and testing at higher emitter temperatures, could result in substantially higher efficiencies than those obtained with our single-junction 0.74-eV GaInAs device.

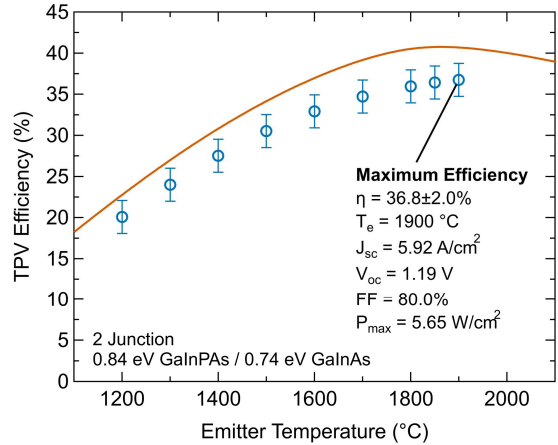

**Figure 6. Two-junction TPV efficiency.**
Measured (points) and modeled (solid line) heat-to-electricity conversion efficiency of a two-junction 0.84-eV GaInPAs / 0.74-eV GaInAs cell with 0.8 cm$^2$ total and 0.64 cm$^2$ aperture areas. Higher voltages and lower current densities allow series resistance to play a less significant role compared to a single-junction cell, which is why the efficiency is still rising at the maximum measured temperature of 1900 °C.

## CONCLUSIONS

We demonstrate large-area and scalable single-junction and tandem TPV cells that achieve high efficiencies and power densities with emitter temperatures approaching 2000°C. The 0.74-eV GaInAs device reaches an efficiency of 38.8±2.0% and power density of 3.78 W/cm$^2$ with an emitter temperature of 1850°C, and the 0.84-eV GaInPAs / 0.74-eV GaInAs device reaches an efficiency of 36.8±2.0% and power density of 5.65 W/cm$^2$ at an emitter temperature of 1900°C. These high efficiencies, especially for the single-junction cell, are enabled by optimizing both the spectral management and charge collection characteristics of the device. Spectral management efficiency benefits from operating with a high emitter temperature, which in turn is made possible by having low series resistance to improve charge collection efficiency. The preliminary tandem device shows promise for reaching even higher efficiencies due to its ability to operate with lower current densities and more efficiently utilize the radiation spectrum of very high temperature thermal emitters.

Both types of devices are designed to be scalable and transferrable to commercial epitaxy systems. For this reason, all layers of both devices are grown lattice-matched to InP substrates, which provides excellent material quality and does not require complex metamorphic buffers to change the lattice constant of the material. The single-junction cell is shown to have repeatable, excellent electrical characteristics across devices on a single wafer and across many growths on different wafers, which further demonstrates the scalability of this cell structure. These designs therefore have the potential to serve as a basis for TPV devices in a variety of high-temperature applications.

## METHODS

### Efficiency Metrics

To reach high efficiencies, TPV systems must minimize thermal, electrical, and optical losses. Thermal losses occur primarily at the system level and include conversion or transfer losses from the heat source to the thermal emitter (described by the source transfer efficiency $TE$) as well as convection and cavity losses of heat from the emitter (described by the cavity efficiency $CE$).[7] These can be minimized through system engineering and scaling to larger sizes. Accordingly, we can describe the system efficiency as

$$\eta_{system} = TE \cdot CE \cdot \eta_{TPV} \tag{3}$$

Electrical and optical losses are specific to the TPV emitter-cell pair and affect the TPV efficiency given in Eq. 1 and repeated here for convenience:

$$\eta_{TPV} = \underbrace{(SE \cdot IQE)}_{\text{Spectral Management Efficiency}} \cdot \underbrace{(VF \cdot FF)}_{\text{Charge Collection Efficiency}} \tag{4}$$

The product of spectral efficiency $SE$ and internal quantum efficiency $IQE$ is[6]

$$SE \cdot IQE = \frac{E_g \int_{E_g}^{\infty} \varepsilon_{eff}(E) \cdot IQE \cdot b(E, T_e) dE}{\int_0^{\infty} \varepsilon_{eff}(E) \cdot E \cdot b(E, T_e) dE} \tag{5}$$

where $E_g$ is the bandgap energy, $\varepsilon_{eff}(E) = \frac{\varepsilon_e \varepsilon_c}{\varepsilon_e + \varepsilon_c - \varepsilon_e \varepsilon_c}$ is the effective emittance of the cavity formed by the emitter with emittance $\varepsilon_e$ and cell with emittance $\varepsilon_c$, $b(E, T_e) = \frac{2\pi E^2}{c^2 h^3 \left[\exp\left(\frac{E}{kT}\right) - 1\right]}$ is the spectral photon flux from a blackbody, $E$ is the photon energy, and $T_e$ is the emitter temperature. The voltage factor $VF$ is

$$VF = \frac{qV_{oc}}{E_g} \tag{6}$$

where $V_{oc}$ is the open-circuit voltage and $q$ is the charge of an electron. The fill factor $FF$ is

$$FF = \frac{J_{mpp} V_{mpp}}{J_{sc} V_{oc}} \tag{7}$$

where $J_{mpp}$ and $V_{mpp}$ are the current density and voltage at the maximum power point and $J_{sc}$ is the short-circuit current density. The radiative limits of all these quantities are provided by Burger et al.[6]

### Cell Growth and Fabrication

Cells were grown by atmospheric pressure metalorganic vapor phase epitaxy in a custom-built cold-wall reactor. Source gases included trimethylgallium, trimethylindium and trimethylaluminum for the group-III elements, arsine and phosphine for the group-V elements, and dilute hydrogen selenide, disilane, diethylzinc and carbon tetrachloride for the dopants. All source gases were mixed into a purified hydrogen carrier gas flowing at 6 lpm.

The substrate was a sulfur-doped (001) InP wafer, miscut 2° toward the (111) direction. The epi-ready substrate was loaded onto a graphite susceptor in the middle of the reactor and heated under a phosphine overpressure to 620°C over the course of five minutes, followed by a 1 minute deoxidation at 620°C and then growth of a ~0.1 μm InP seed layer, and then a layer each of GaInAs and InP etch-stops for post-growth processing. For the single-junction cells, the remainder of the layers in Fig. 2A were grown at 620°C, at growth rates of 6.8 μm/hour for the arsenide layers and 3.7 μm/hour for the InP layers. The composition of the GaInAs layers is $Ga_{0.47}In_{0.53}As$ determined by Vegard's law for GaInAs lattice-matched to InP.[52] The sample was cooled to room temperature under an arsine overpressure.

For the two-junction samples shown in Fig. S4, the GaInPAs top cell was grown at 665°C, with ~0.5 μm GaInPAs replacing the GaInAs absorber in Figure 2A. The composition of the GaInPAs layer is $Ga_{0.35}In_{0.65}P_{0.25}As_{0.75}$ determined from the calculation method of Moon *et al*.[53] After the growth of the InP back surface field layer, a thin GaInAs buffer layer was grown and then the reactor was cooled to 550°C. The tunnel junction consisted of ~30 nm carbon-doped GaAsSb and ~25 nm selenium-doped InP. The reactor was then heated back to 620°C for the growth of the GaInAs bottom cell as described above.

Fabrication of individual cells was completed with standard photolithographic, wet etching, and electroplating processes. The Au rear reflector was first deposited by electroplating on the epitaxial surface. The samples were then bonded, Au-side down, to an intrinsic Si wafer with thermally-conductive epoxy in order to provide mechanical stability.[54] The InP substrate was chemically etched away in $HCl : H_3PO_4$ (4:1), and the GaInAs and InP etch-stop layers were removed with $H_3PO_4 : H_2O_2 : H_2O$ (3:4:1) and HCl, respectively. The front metal (thin Ni adhesion layer and Au) was photolithographically patterned using Megaposit SPR220-4.5 photoresist and deposited *via* electroplating. Finally, separate cell areas were isolated by protection with patterned photoresist and chemical etching with the previously mentioned etchants to expose the rear reflector in between cells. Individual cells were cleaved from the parent substrate for TPV testing.

**Cell Characterization**

Quantum efficiency was measured on a custom-built instrument consisting of a tungsten-halogen lamp and a 270M monochromator. The incident light was chopped at 313 Hz. QE was measured every 5 nm from 350-1850 nm. The incident flux at each wavelength was measured with a calibrated GaInAs cell and a beam splitter. The output current at each wavelength was measured with a low-noise current preamplifier and a lock-in amplifier. Specular reflectance was measured simultaneously, using a calibrated stacked Si/Ge photodiode and a separate pre-amplifier and lock-in amplifier. For the two-junction cells in Figure S5, high brightness Mightex LEDs at 656 nm and 1550 nm were used to separate the responses of the individual sub-cells. The effects of luminescent coupling were removed using the methods of Steiner *et al*.[55]

Current-voltage curves were measured at "one-sun" on an XT-10 solar simulator, using a xenon light source. The cells were measured under a photon flux that simulated the AM1.5 G173-03 direct solar spectrum[56] at 1000 W/m², with the intensity set using a calibrated GaInAs reference cell and a spectral mismatch correction.[57] The incident spectrum was measured with a spectral evolution spectrophotometer. We used a Keithley 238 low noise source-meter to measure the current at each voltage, sweeping from forward to reverse bias in 5 mV steps.

Directional-hemispherical reflectance of the cells was measured with a Bruker Invenio-X spectrometer equipped with a PIKE Technologies IntegratIR diffuse gold integrating sphere with a 12° angle of incidence using the substitution method.[58] All measurements were referenced to a NIST standard with known reflectance ±0.3%. Sample-to-sample deviations have resulted in an uncertainty of 0.4%, which results in a ±0.5% absolute total uncertainty.

**TPV Efficiency Testing**

We measure TPV efficiency using a calorimetric approach. As seen in the schematic in Figure 7, we place the cell on a cutbar, which is a metal post with regularly spaced thermistors. The cutbar is mounted on a temperature-controlled heatsink. The heat flowing through the cutbar can be calculated using the formula

$$\dot{Q} = k\frac{A}{l}\Delta T \quad (8)$$

where $k$ is the thermal conductivity, $A$ is the area of the cross section of the cutbar, $l$ is the spacing between thermistors, and $\Delta T$ is the difference between the thermistor readings. The heat conducted through the electrical probes can be quantified similarly, as we place thermocouples on the probes.

We illuminate the TPV cell with a thermal emitter through a diamond window resting on a copper aperture. The water-cooled aperture protects the electronics from the intense thermal radiation and the diamond window prevents sublimed carbon from the emitter from depositing on the cell. The thermal emitter is surrounded by carbon-based insulation to prevent excessive heating of the apparatus. A small hole in the insulation allows an optical pyrometer to monitor the temperature of the thermal emitter. We pass current through graphite strips that radiatively heat the emitter. The entire assembly is inside a bell jar that is back-filled with argon to prevent oxidation.

We monitor maximum power point (MPP) of the cell as well as the heat passing through the cutbar and electrical probes throughout the experiment. After increasing or decreasing the temperature, we wait until the MPP has stabilized and then collect an current-voltage sweep. The efficiency is simply the ratio of the electrical power at MPP to the sum of electrical power and heat.

We also quantify the view factor between the cell and emitter to corroborate the measured efficiency with our model. Using a home-built Monte Carlo ray tracing script, we track the number of photons that reach the cell from the emitter and use that to approximate the view factor. We also place a nearly 100% absorptive metal velvet strip on the cutbar and determine the power passing through the cutbar as we change the emitter temperature. Both methods agree that the view factor is approximately 31%.

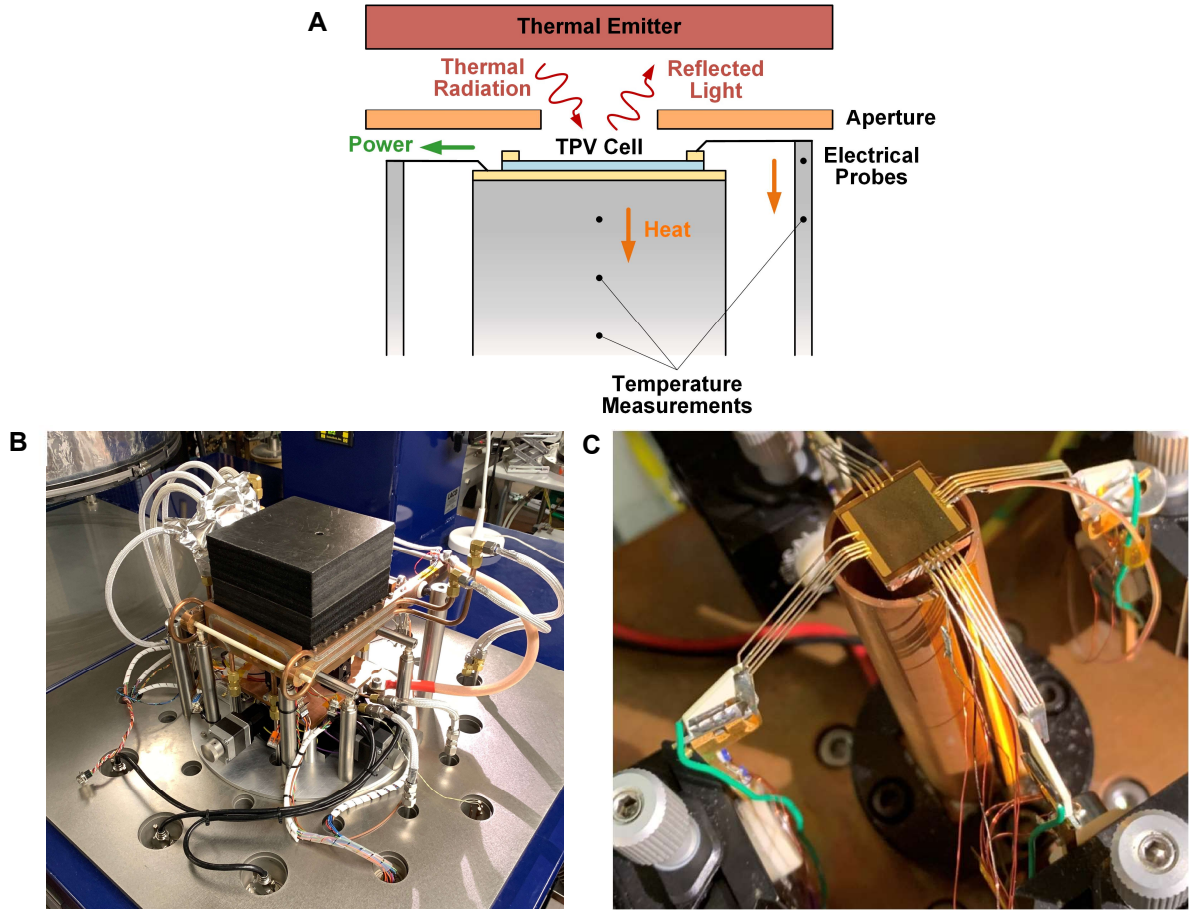

**Figure 7. Schematic and photos of TPV measurement facility.**
(A) Test facility schematic. Thermal radiation from a heated carbon emitter is incident on the TPV cell through a water-cooled copper aperture, and most sub-bandgap light is reflected back to the emitter. The cell is mounted on a metal pedestal, and electrical probes make contact with the rear reflector and the front busbars to measure electrical power produced. Multiple temperature measurements of the pedestal and electrical probes allow calculation of the heat dissipated by the cell.
(B) Photograph of the test facility with insulation and cooling infrastructure in place.
(C) Photograph of a cell mounted in the test facility, showing a radiation shield around the metal cutbar and electrical probes contacting the cell.

**Optoelectronic and Efficiency Modeling**

The series resistance of 6.5 mΩ cm$^2$ for the single junction device was determined by fitting a generalized optoelectronic model[42] to the open-circuit voltage, $FF$, and short-circuit current density data in Fig. 3C as well as to separate electroluminescence measurements that provided the external radiative efficiency and dark current-voltage characteristics, shown in Fig. S7. For the electroluminescence measurements, a calibrated spectroradiometer was used to collect the emitted light from the forward-biased cell as in references [42, 59].

The spectral absorptance and reflectance of the cells was calculated with the standard transfer matrix method.[45] For this model, the optical properties of the semiconductor layers were determined by ellipsometry. The effects of doping were accounted for by adding the imaginary part of the relative permittivity from a Drude model for free-carrier absorption.[60] Optical properties of the gold front metal and back reflector were taken from reference [61].

To model the performance of our TPV cells, we re-write Eqn. 2 as

$$\eta_{TPV} = \frac{P}{P+Q} = \frac{V_{oc}J_{sc}FF}{Q_{inc} - Q_{ref}} \qquad (9)$$

where $Q_{inc}$ is the incident radiation power density on the TPV cell and $Q_{ref}$ is the reflected radiation. To model $V_{oc}$, $J_{sc}$, and $FF$, we used the same well-established optoelectronic model[42] that was used to determine series resistance of the single-junction cell. Inputs to the model include the bandgap-$V_{oc}$ difference at one sun, which was measured experimentally (Fig. S1), and the series resistance. Within the model, $J_{sc}$ was calculated by assuming a blackbody emitter with a view factor to the cell as described above and the measured external quantum efficiency:

$$J_{sc} = \frac{q}{c \cdot h} VF \int_0^\infty EQE(\lambda) Q_{bb}(\lambda, T) \lambda \, d\lambda \qquad (10)$$

where $Q_{bb}(\lambda, T)$ is the spectral blackbody emissive power. To account for variations in cell temperature (Fig. S2), the model considers the effect of temperature on the intrinsic carrier concentration and the dark current as well as its effect on the shift in bandgap energy.[62, 63] For the single-junction cell, the previously-determined series resistance was utilized with the model. For the two-junction cell, the series resistance was used as a fitting parameter and determined to be 10 mΩ cm² (Fig. S5C). As seen in Fig. 8, the model provides good agreement with experimental data.

The performance characteristics shown in Fig. 8 are all readily understandable consequences of photovoltaic device physics, as explained in the discussion of Fig. 3 above. The one trend needing further explanation here is the fill factor of the two-junction device. Superposed on the overall series-resistance-driven decrease in the fill factor with increasing emitter temperature, the two-junction fill factor goes through a minimum at ~1700-1800°C. This minimum is driven by the ratio of the top to the bottom junction photocurrents ($J_{sc,top}/J_{sc,bot}$). At the lower emitter temperatures, $J_{sc,top}/J_{sc,bot} < 1$, but $J_{sc,top}/J_{sc,bot}$ increases with emitter temperature as the emitted spectral content shifts to higher photon energies, which causes $J_{sc,top}$ and $J_{sc,bot}$ to go through a current-matched point $J_{sc,top}/J_{sc,bot} = 1$ at ~1700-1800°C. The fill factor is a minimum at the current matching condition, a general feature of series-connected two-junction PV cells.[64]

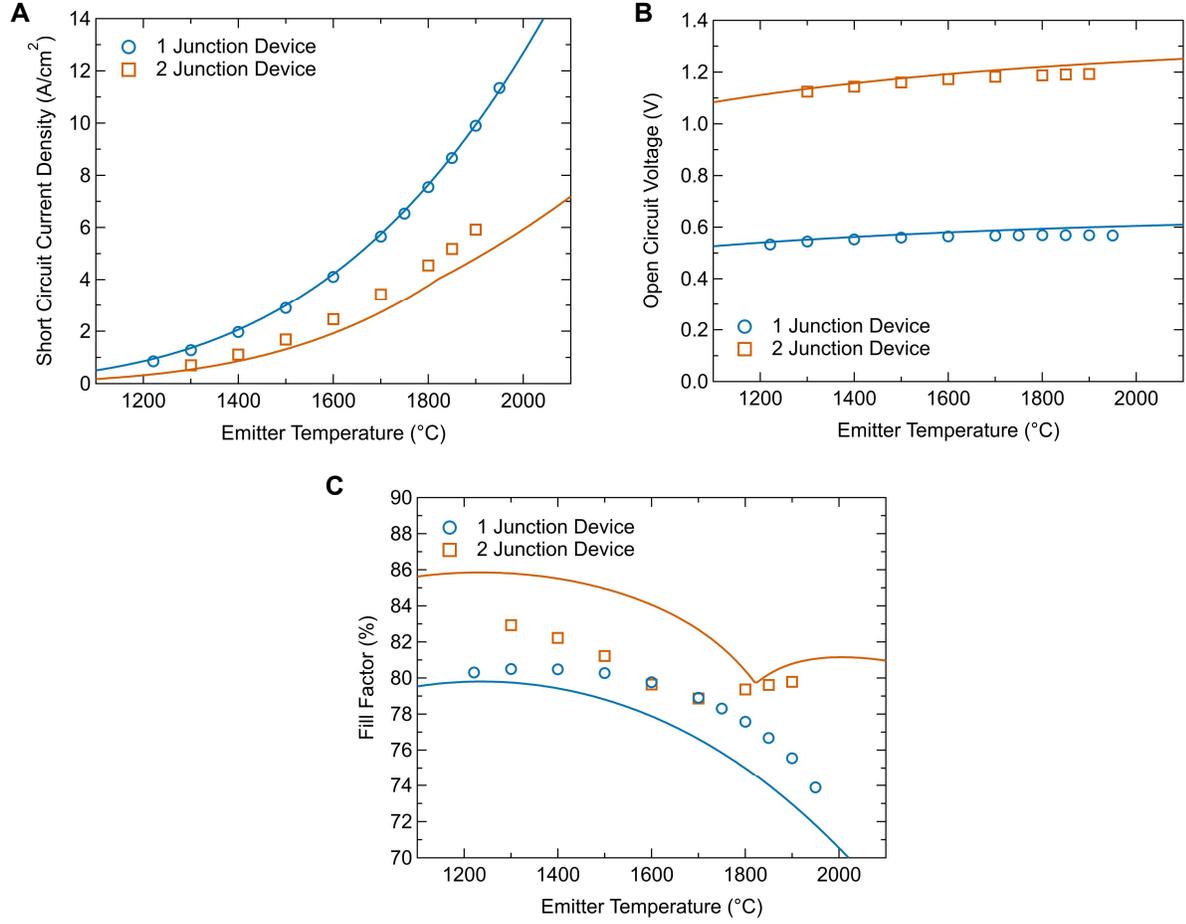

**Figure 8. Electrical characteristics during TPV testing.**
Measured (points) and modeled (lines) electrical characteristics of single- and two-junction cells during TPV efficiency testing. Data includes:
(A) Short circuit current density
(B) Open circuit voltage
(C) Fill factor

The incident power is calculated assuming blackbody emission with a view factor to the cell as described above:

$$P_{inc} = VF \int_0^\infty Q_{bb}(\lambda, T) \mathrm{d}\lambda \qquad (11)$$

The reflected power is calculated similarly but also using the measured spectral directional-hemispherical reflectance $\rho(\lambda)$ (Fig. 4A and Fig. S6A):

$$P_{ref} = VF \int_0^\infty \rho(\lambda) Q_{bb}(\lambda, T) \mathrm{d}\lambda \qquad (12)$$

ACKNOWLEDGEMENTS

This work was authored in part by the National Renewable Energy Laboratory, operated by Alliance for Sustainable Energy, LLC, for the U.S. Department of Energy (DOE) under Contract No. DE-AC36-08GO28308. Funding was provided by the U.S. Department of Energy Advanced Research Projects Agency-Energy under cooperative agreement DE-AR0000993 and by the Shell Gamechangers program under contract number ACT-18-36. The views expressed in the article do not necessarily represent the views of the DOE or the U.S. Government. The U.S. Government retains and the publisher, by accepting the article for publication, acknowledges that the U.S. Government retains a nonexclusive, paid-up, irrevocable, worldwide license to publish or reproduce the published form of this work, or allow others to do so, for U.S. Government purposes.

31. Gessert, T.A. and Coutts, T.J. (1992). Grid metallization and antireflection coating optimization for concentrator and one-sun photovoltaic solar cells. Journal of Vacuum Science & Technology A *10*, 2013-2024.
32. Schulte, K.L., France, R.M., Friedman, D.J., LaPotin, A.D., Henry, A. and Steiner, M.A. (2020). Inverted metamorphic AlGaInAs/GaInAs tandem thermophotovoltaic cell designed for thermal energy grid storage application. J. Appl. Phys. *128*, 143103.
33. LaPotin, A., Schulte, K.L., Steiner, M.A., Buznitsky, K., Kelsall, C.C., Friedman, D.J., Tervo, E.J., France, R.M., Young, M.R., Rohskopf, A., et al. (2022). Thermophotovoltaic efficiency of 40%. Nature *604*, 287-291.
34. Geisz, J.F., Kurtz, S., Wanlass, M.W., Ward, J.S., Duda, A., Friedman, D.J., Olson, J.M., McMahon, W.E., Moriarty, T.E. and Kiehl, J.T. (2007). High-efficiency GaInP∕GaAs∕InGaAs triple-junction solar cells grown inverted with a metamorphic bottom junction. Appl. Phys. Lett. *91*, 023502.
35. Steiner, M.A., Geisz, J.F., García, I., Friedman, D.J., Duda, A. and Kurtz, S.R. (2013). Optical enhancement of the open-circuit voltage in high quality GaAs solar cells. J. Appl. Phys. *113*, 123109.
36. Yablonovitch, E., Gmitter, T., Harbison, J.P. and Bhat, R. (1987). Extreme selectivity in the lift-off of epitaxial GaAs films. Appl. Phys. Lett. *51*, 2222-2224.
37. Youtsey, C., Adams, J., Chan, R., Elarde, V., Hillier, G., Osowski, M., McCallum, D., Miyamoto, H., Pan, N., Stender, C., et al. (2012). Proc. of International Conference on Compound Semiconductor Manufacturing Technology Boston, MA.
38. Sweet, C.A., Schulte, K.L., Simon, J.D., Steiner, M.A., Jain, N., Young, D.L., Ptak, A.J. and Packard, C.E. (2016). Controlled exfoliation of (100) GaAs-based devices by spalling fracture. Appl. Phys. Lett. *108*, 011906.
39. Chen, J. and Packard, C.E. (2021). Controlled spalling-based mechanical substrate exfoliation for III-V solar cells: A review. Sol. Energy Mater. Sol. Cells *225*, 111018.
40. Coll, P.G., Neumann, A., Smith, D., Warren, E., Polly, S., Hubbard, S., Steiner, M.A. and Bertoni, M.I. (2021). Proc. of 2021 IEEE 48th Photovoltaic Specialists Conference (PVSC), DOI: 10.1109/PVSC43889.2021.95186562141-2143.
41. Steiner, M.A., Perl, E.E., Geisz, J.F., Friedman, D.J., Jain, N., Levi, D. and Horner, G. (2017). Apparent bandgap shift in the internal quantum efficiency for solar cells with back reflectors. J. Appl. Phys. *121*, 164501.
42. Geisz, J.F., Steiner, M.A., García, I., France, R.M., McMahon, W.E., Osterwald, C.R. and Friedman, D.J. (2015). Generalized Optoelectronic Model of Series-Connected Multijunction Solar Cells. IEEE Journal of Photovoltaics *5*, 1827-1839.
43. King, R.R., Law, D.C., Edmondson, K.M., Fetzer, C.M., Kinsey, G.S., Yoon, H., Sherif, R.A. and Karam, N.H. (2007). 40% efficient metamorphic GaInP∕GaInAs∕Ge multijunction solar cells. Appl. Phys. Lett. *90*, 183516.
44. King, R.R., Bhusari, D., Boca, A., Larrabee, D., Liu, X.Q., Hong, W., Fetzer, C.M., Law, D.C. and Karam, N.H. (2011). Band gap-voltage offset and energy production in next-generation multijunction solar cells. Progress in Photovoltaics: Research and Applications *19*, 797-812.
45. Yeh, P. (2005). Optical Waves in Layered Media (John Wiley & Sons, Inc.).
46. Burger, T., Fan, D., Lee, K., Forrest, S.R. and Lenert, A. (2018). Thin-Film Architectures with High Spectral Selectivity for Thermophotovoltaic Cells. ACS Photonics *5*, 2748-2754.
21

# SUPPLEMENTARY INFORMATION

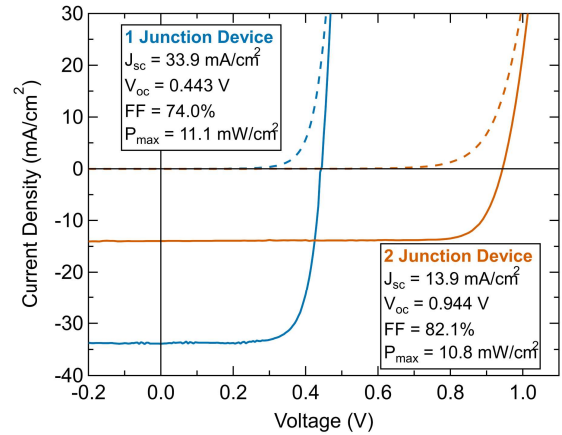

**Figure S1. One-sun current-voltage characteristics of TPV cells.**
The measured current-voltage characteristics of single- and two-junction cells under dark conditions (dashed lines) and AM1.5D illumination (solid lines). The single-junction device in particular exhibits an excellent open-circuit voltage only 0.297 V less than the bandgap voltage.

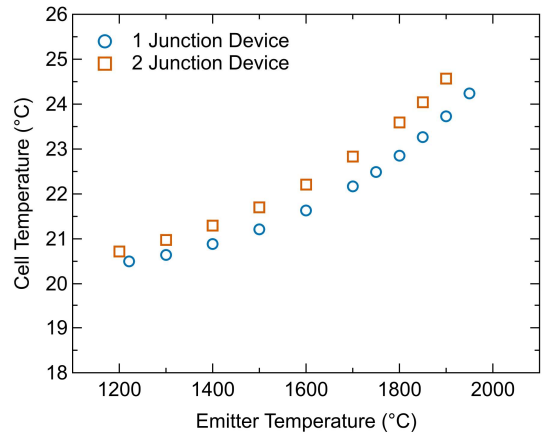

**Figure S2. TPV cell temperatures during testing.**
As the temperature of the emitter is increased, the cell temperature also increases slightly due to limited capabilities of the cooling system.

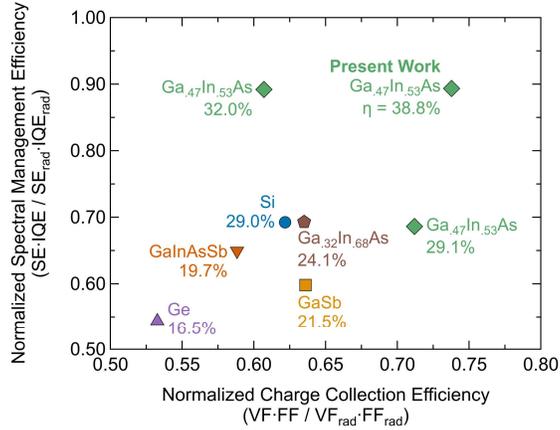

**Figure S3. Normalized efficiencies of TPV devices.**
Historical and present experimental single-junction TPV efficiencies (indicated by point labels) plotted against the spectral management and charge collection efficiencies normalized to the radiative limit.

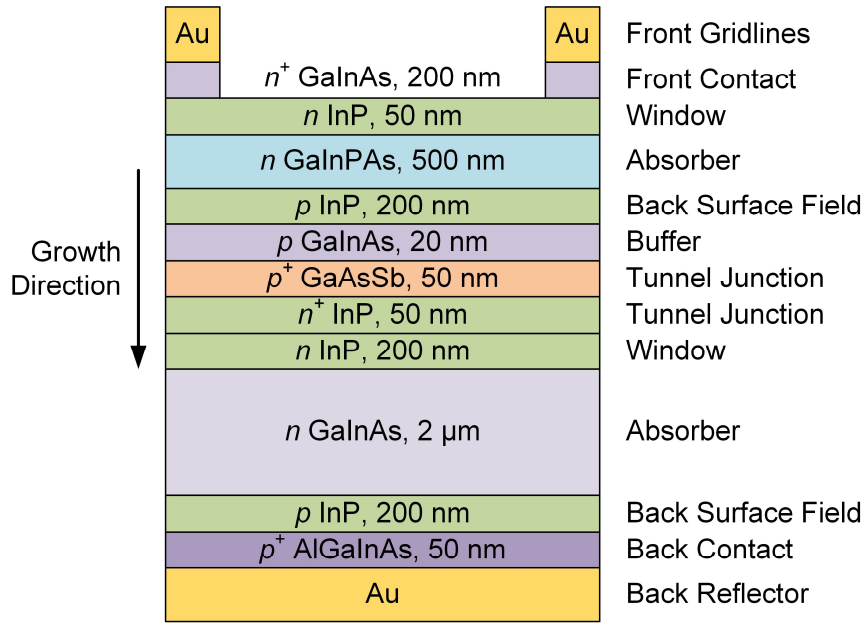

**Figure S4. Two-junction GaInPAs / GaInAs TPV cell.**
Schematic of the tandem TPV cell structure. The tunnel junction layers are kept as thin and low-doped as possible to reduce parasitic above- and below-bandgap absorption.

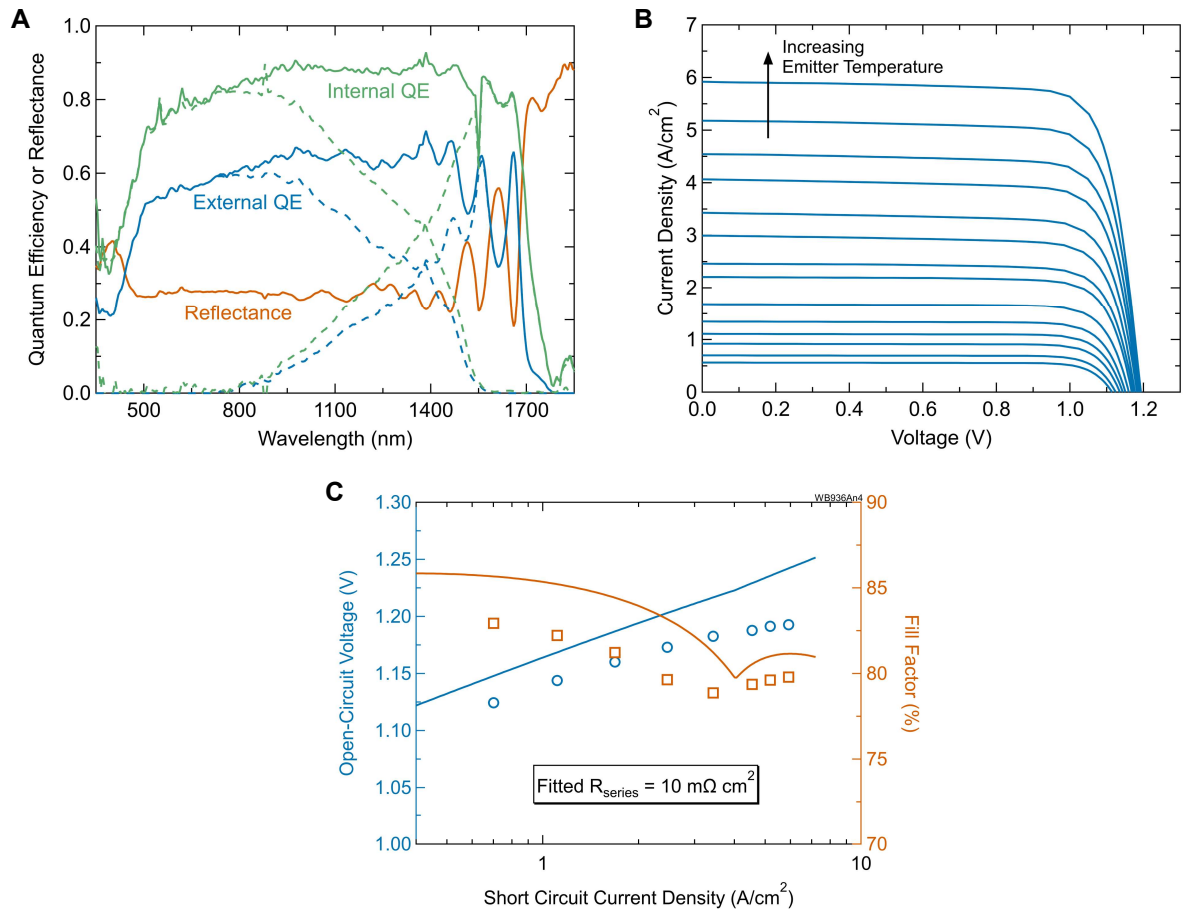

**Figure S5. Two-junction TPV cell electrical characteristics.**
(A) Measured reflectance and external quantum efficiency and calculated internal quantum efficiency for a two-junction GaInPAs / GaInAs TPV cell. Dashed lines are for individual junctions and solid lines are the summation of individual junctions.
(B) Series of current-voltage measurements of a two-junction TPV cell under an increasing emitter temperature.
(C) Open-circuit voltage (blue circles) and fill factor (orange squares), the latter from a two-diode model fit, extracted from (B). Solid lines are data fit to a generalized optoelectronic model that provide the series resistance.

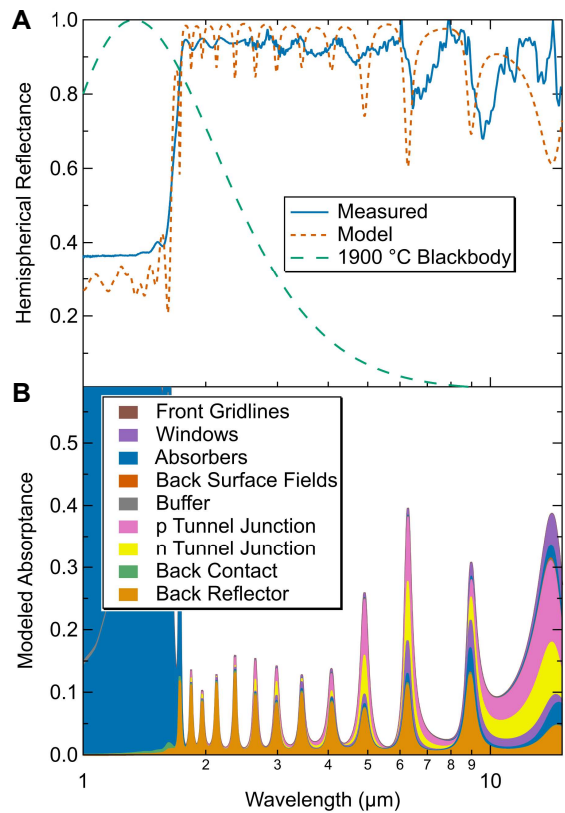

**Figure S6. Two-junction TPV cell optical characteristics.**
(A) Measured and modeled hemispherical reflectance of tandem TPV cell. Also shown is the normalized 1900 °C blackbody spectrum, which indicates that the amount of sub-bandgap energy at wavelengths longer than 15 μm is negligible.
(B) Modeled absorptance in each layer for the same case as (A). Parasitic absorption for wavelengths longer than the bandgap occurs primarily in the Au back reflector and the tunnel junction layers.

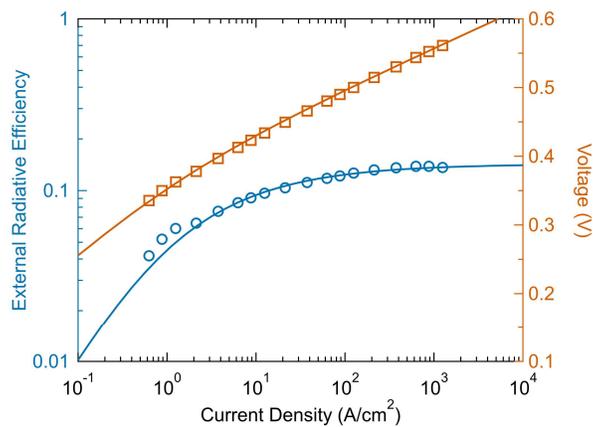

**Figure S7. Single-junction electroluminescence measurements and fit.**
External radiative efficiency (blue circles) and voltage (orange squares) as a function of injected current. Solid lines are fits of the generalized optoelectronic model described in the Methods.

27